\documentstyle[12pt]{article}

\newcommand{\be}{\begin{equation}}
\newcommand{\ee}{\end{equation}}
\newcommand{\bea}{\begin{eqnarray}}
\newcommand{\eea}{\end{eqnarray}}

\begin{document}

\begin{center}
\begin{large}
{\bf  dS/CFT Duality on the Brane \\}
{\bf with a\\}
{\bf  Topological Twist \\}
\end{large}  
\end{center}
\vspace*{0.50cm}
\begin{center}
{\sl by\\}
\vspace*{1.00cm}
{\bf A.J.M. Medved\\}
\vspace*{1.00cm}
{\sl
Department of Physics and Theoretical Physics Institute\\
University of Alberta\\
Edmonton, Canada T6G-2J1\\
{[e-mail: amedved@phys.ualberta.ca]}}\\
\end{center}
\bigskip\noindent
\begin{center}
\begin{large}
{\bf
ABSTRACT
}
\end{large}
\end{center}
\vspace*{0.50cm}
\par
\noindent

We consider a brane universe in an  asymptotically  de Sitter
 background spacetime  of arbitrary dimensionality.
In particular, the bulk spacetime  is described by a
``topological de Sitter'' solution, which has recently been  investigated
by Cai, Myung and Zhang.  
 In the current study,  we begin by  showing
that the brane evolution is described by  Friedmann-like
equations  for radiative matter.
 Next, on the basis of the  dS/CFT correspondence,
we identify the thermodynamic properties of the brane universe.
We then demonstrate that many (if not all) of the holographic aspects
of  analogous  AdS-bulk scenarios persist.
These include a (generalized) Cardy-Verlinde form for the CFT
entropy  and various coincidences when the brane crosses the
cosmological horizon.

\newpage

\section{Introduction}
\par

It is not uncommon to find the same physical system being
described by two or more  seemingly  unrelated pictures.
Nowhere is this ambiguity more apparent than
in recent  attempts at describing the  universe itself. 
For instance, we  have seen 11-dimensional membrane theory
 give
rise to an abundance of dualities when its various manifestations
are appropriately compactified \cite{duff}. On the other hand,
we have it, on pretty good authority, that the physical universe
can be effectively described by merely four spacetime dimensions.
Such  duality between theories of  distinct  dimensionality
may, in fact, be a consequence of a more fundamental 
concept; namely, 
the  ``holographic principle'' \cite{tho,sus}. 
\par  
The underlying premise of ``holography''  is that the maximal entropy
within any given volume will  be determined by the largest black
hole that fits inside of that volume \cite{book}.  Since  the entropy
of a black hole is
 (up to  a constant factor) given by its horizon  surface area 
\cite{bek,haw}, 
it follows   
that the relevant degrees of freedom of a black hole must,
 in some sense, ``live'' on  the horizon.  Moreover,
 given the holographic premise,
 it follows that the relevant degrees of {\bf any} system
must live on a surface that bounds   the volume of
that system.
\par
 The holographic principle has played its (perhaps) most
prominent role in establishing a  duality that seems to exist
between
any  anti-de Sitter (AdS) spacetime\footnote{Note that
 anti-de Sitter denotes   gravity  with a negative
cosmological constant and de Sitter, a positive
cosmological constant.  Typically, the
gravity is described by Einstein theory, but not exclusively so.}
 and a lower-dimensional
 conformal field theory (CFT) \cite{mal,gub,wit}.
More specifically,  it has been  convincingly argued  that 
 the horizon thermodynamics  of an  $n$+2-dimensional AdS
 black hole can be  identified with  a certain
 $n$+1-dimensional (strongly-coupled)
CFT.  Significantly to these arguments,  the dual CFT is assumed
 to live on a
timelike surface that can be identified as an asymptotic boundary
of the AdS spacetime.
\par
In analogy to this well-accepted AdS/CFT duality, a 
de Sitter(dS)/CFT
correspondence  has  similarly been conjectured \cite{str}. (For
earlier works in this regard, see Refs.\cite{first3}-\cite{last3}.)
Although dS space is obtained from AdS  with a seemingly trivial
sign change in the cosmological constant, there turns out to
be quite  severe  implications. As a consequence,  
 one finds that establishing the dS/CFT duality is a much more
difficult challenge than in the AdS case.
 For example,  dS space lacks a 
 globally timelike Killing vector and a spatial infinity (making
it difficult to define conserved charges), while
  the black hole horizon (and its  
thermodynamic properties)
 have an
ambiguous observer
dependence.\footnote{For a comprehensive discussion on
dS spacetimes, see Ref.\cite{str2}.}  It is also problematic
 that dS solutions are conspicuously absent in  string theories
(and  other  quantum gravity theories); thus  
impeding any rigorous  testing of the proposed duality.
\par
In spite of these inherent  complications,  
there has still been significant progress 
towards a holographic understanding of dS spacetimes 
\cite{str}-\cite{mann}.  
    With regard to the conjectured  
correspondence, the 
  dS cosmological horizon is used in place of the (inner-lying)
black hole horizon. Furthermore, 
the dual CFT is regarded as a Euclidean one that lives on a spacelike
asymptotic boundary.  
Essential to these identifications 
 is  a renormalization group  flow 
 (between Euclidean CFTs
at past and future infinity) that happens to be dual with time evolution
in the dS bulk \cite{str3}. 
\par
Let us return  our attentions, for the moment, to
the AdS/CFT correspondence.  
In a  relevant paper \cite{ver}, Verlinde  directly applied
this holographic   duality to 
a  radiation-dominated 
Friedmann-Robertson-Walker (FRW) universe 
(in $n$+1 dimensions).\footnote{For earlier
studies on holography in a cosmological setting, 
see Refs.\cite{first}-\cite{last}.} 
This paper had a wide scope, but  two observations are 
of particular  interest. (i)  The AdS/CFT correspondence
 leads to a CFT  entropy that can be
expressed in terms of a generalized Cardy formula \cite{car}; with  
 the Cardy ``central charge''  being a direct manifestation of the Casimir
energy.\footnote{In this context, the Casimir energy
refers specifically to the sub-extensive portion of
the thermal energy. Furthermore, we will refer to the corresponding entropy
as the ``Casimir entropy''.} (ii) When the Casimir entropy  
 saturates a certain bound (namely,
 the  Bekenstein-Hawking entropy \cite{bek,haw}
 of  a universal-size
black hole), then  the cosmological evolution or Friedmann equations 
coincide 
with the generalized Cardy formula. We can express this point more eloquently:
the CFT and FRW equations  merge at a  holographic
saturation point, which implies  that both sets of
equations arise from some fundamental, underlying theory.
\par
In Ref.\cite{sav}, Savonije and Verlinde  have extended the prior work
 to an intriguing scenario: a  Randall-Sundrum brane world \cite{ran,cos}
in the background of an AdS-Schwarzschild (black hole) geometry. 
In this context,
the $n$+1-dimensional CFT is regarded as living on the brane, which serves
as a suitable  boundary for the $n$+2-dimensional AdS bulk spacetime.
With  appropriately chosen  boundary conditions,  Savonije and Verlinde have
 shown
that the brane world corresponds to a  FRW universe and the brane dynamics
are  described by the Friedmann equations for radiative matter.
Moreover, it was shown that the CFT thermodynamic relations
coincide with the Friedmann  equations  at a special cosmological point:
 when the
brane intersects the black hole  horizon. 
\par
Here, we note that many aspects of the  Verlinde-Savonije  program   have
since  been extended and generalized. The relevant studies
(for an AdS scenario)
can be found in Refs.\cite{first2}-\cite{last2}.
\par
Most  recently,  the  Verlinde-Savonije
treatment  \cite{ver,sav} has been extended to a  
 dS/CFT holographic picture  
\cite{new1}-\cite{med}.\footnote{Such studies
may be of particular importance, given
recent empirical evidence
of  a positive
cosmological constant for our universe
 \cite{obs}.}
These studies were, for the most part, successful in generalizing
the pertinent features of Refs.\cite{ver,sav}  to   dS scenarios.
However, there were some bothersome issues that can be directly attributed
to the inherent complexities of dS spacetimes. 
These issues  include negative
energy densities on the CFT boundary (also see Refs.\cite{baln,myun}),
the total  CFT  entropy being bounded from above by the Casimir
contribution (also see Ref.\cite{booz}), 
 the CFT-based  universe being inaccessible  
 to a strongly self-gravitating regime (especially see Ref.\cite{med}),
and an  inability to incorporate  the thermodynamics of
the relevant  black hole horizons 
into the proposed  duality.
\par
A preliminary analysis by Cai \cite{new3} suggests that many
(if not all) of these issues can be resolved by revising 
the duality to incorporate  a certain brand
of asymptotically dS geometries. These ``topological de Sitter'' (TdS)
spacetimes were originally proposed 
in Ref.\cite{last4}. However,  there is a ``cost'' to be extracted
if one is to proceed along these lines. Such TdS spacetimes
have no black hole horizon,  and  so a naked singularity is
an inevitable consequence.\footnote{In fact, the original motivation
 for considering TdS spacetimes \cite{last4} was to test an earlier
 conjecture
on cosmological singularities \cite{bal}.}
  On the other hand,  the existence
of  a well-defined
CFT  that can describe this  singularity does  not 
seem inconceivable.
Given this possibility, it seems worth pursuing if
the pertinent outcomes of Ref.\cite{sav} hold up under a TdS-bulk
 scenario. Just such an investigation is the focus of the current paper.
\par
The  rest of the paper is organized  as follows.  In Section 2,
we identify the thermodynamics properties of the TdS cosmological
horizon. We also formulate   the brane dynamics,
which are shown to be described  by Friedmann-like equations.
In Section 3,  we apply the   dS/CFT correspondence and
identify 
 the brane (or CFT) thermodynamic properties.  
 Also,  the generalized Friedmann equations are
re-expressed  so that their connection with CFT thermodynamics
is manifest.
In Section 4, we  
 demonstrate
that the CFT thermodynamic  and  Friedmann equations coincide
when the brane crosses the horizon.  In addition, the CFT entropy
is shown to  be expressible in a  Cardy-Verlinde-like form \cite{car,ver}.
 Section 5   considers the  
 holographic entropy bounds in the context of this model. 
Finally, Section 6 ends with
a summary and brief discussion.

\section{Bulk Thermodynamics and Brane Cosmology}
\par
We  begin the analysis by formulating  the
scenario of interest. Namely, a $n$+1-dimensional brane of constant tension
in an $n$+2-dimensional  topological de Sitter (TdS) background.
In a suitably static gauge, the bulk solution can be written
as follows \cite{last4,new3}
\be
ds^2_{n+2}=-h(a)dt^2+{1\over h(a)}da^2+a^2d\Omega^2_{n},
\label{1}
\ee
\be
h(a)=k-{a^2\over L^2}+{\omega_{n+1}M\over a^{n-1}},
\label{2}
\ee
\be
\omega_{n+1}={16\pi G_{n+2}\over n V_n}.
\label{3}
\ee
Here, $L$ is the curvature radius of the dS background, $d\Omega_n^2$
denotes the line element of an $n$-dimensional
(constant-curvature) hypersurface
 with volume $V_{n}$, $G_{n+2}$ is
the $n$+2-dimensional Newton constant, and $M$ 
 and $k$  are constants of integration. $M$ is roughly associated
with the mass of the solution\footnote{More precisely, $M$ measures
an excitation in gravitational energy relative 
to the pure ($M=0$) dS spacetime.}
 and will be  regarded as non-negative.
(Note the sign reversal in this term relative to the usual
Schwarzschild-dS case.)
Meanwhile, without loss of generality, $k$ can be set to +1, 0 or -1. 
These choices describe a (cosmological) horizon
geometry that is respectively elliptic, flat or hyperbolic.
 \par
Clearly, the above  solution is asymptotically  dS. However,
the $M\geq 0$ condition  leads to some distinguishing features.
For instance, there is no black hole horizon
(although a cosmological one). Moreover, there exists  a naked
singularity at $a=0$ for any  $M > 0$. However, we will assume that this
singularity can  somehow be described (in a non-singular fashion)
  by the dual CFT of interest and proceed
on this basis. 
\par
For  any asymptotically  dS space, 
there exists a well-defined cosmological horizon having
similar thermodynamic properties to that of a black hole horizon
\cite{str2}. For the above solution, this cosmological horizon ($a=a_H$)  
corresponds to the positive root of $h(a)=0$. Thus, the following
useful relation can be obtained:
\be
k-{a_{H}^2\over L^2} +{\omega_{n+1}M\over a_H^{n-1}}
=0.
\label{5}
\ee
\par
In analogy with black hole thermodynamics \cite{gh}, the cosmological
horizon has an associated temperature and entropy 
that are respectively given as follows:\footnote{In particular, the inverse
temperature can be identified with the periodicity of Euclidean
time and the entropy,  with one quarter of the horizon surface area \cite{gh}.
}
\be
T_{dS}={(n+1)a_{H}^2-(n-1)L^2 k\over 4\pi L^2 a_H} ,
\label{6}
\ee
\be
S_{dS}={a_H^nV_n\over 4 G_{n+2}}.
\label{7}
\ee
The premise of the dS/CFT correspondence is that the above thermodynamics
can be identified, up to a conformal factor, with a Euclidean CFT
that lives  
on a spacelike boundary (for the bulk)
 at temporal infinity. We will re-introduce
and  exploit this duality at an appropriate interval. 
\par
Let us now consider the brane, which can be regarded as a dynamical
boundary of the TdS geometry. To describe these brane dynamics, we
will presume a boundary action of the following form:
\be
{\cal I}_b={1\over 8\pi G_{n+2}}\int _{\partial {\cal M}}
\sqrt{\left| g^{ind}\right|}{\cal K}
+{\sigma\over 8\pi G_{n+2}}\int_{\partial {\cal M}}
\sqrt{\left| g^{ind}\right|},
\label{9}
\ee
where $g^{ind}_{ij}$ is the induced metric on the boundary  
($\partial{\cal  M}$),
${\cal K}\equiv {\cal K}^i_i$ is the trace of the extrinsic curvature
and $\sigma$ is a parameter measuring the brane tension.
By varying this action with respect to the induced metric
(and assuming a one-sided brane scenario),
we
obtain an  equation of motion as follows:
\be
{\cal K}_{ij}={\sigma\over n}g_{ij}^{ind}.
\label{10}
\ee
\par
In analogy with Ref.\cite{sav}, 
we can clarify the brane dynamics by introducing
a new (cosmological) time parameter, $\tau$; whereby  $a=a(\tau)$,
$t=t(\tau)$ and:
\be
{1\over h(a)}\left({da\over d\tau}\right)^2-h(a)\left({dt\over
d\tau}\right)^2=1.
\label{11}
\ee
Unlike in  Ref.\cite{sav},  $\tau$ has been defined
here so as to yield
a spacelike line element.  This choice
naturally reflects the  duality that (presumably) exists between  an
asymptotically  dS spacetime
 and  a Euclidean CFT \cite{str}.
\par
Substituting the above condition into Eq.(\ref{1}), we find
that  the induced brane metric  adopts  a Euclidean FRW form.
More specifically:
\be
ds^2_{n+1}=d\tau^2+a^2(\tau)d\Omega^2_{n}.
\label{12}
\ee
Keep in mind  that the radial distance, $a=a(\tau)$,  
is really just  the size of the
 $n$+1-dimensional brane universe.
\par
Let us now return our attention to  Eq.(\ref{10}); that is,
the boundary equation of motion. One can readily calculate
the extrinsic  curvature (see, for instance, 
Ref.\cite{chr}) and then  express this result  in terms of the functions
$a(\tau)$ and $t(\tau)$. For any of the ``angular
components'' of the induced metric (i.e.,  components
with respect to  the constant-curvature hypersurface),   the described process 
yields:
\be
{dt\over d\tau}={\sigma a\over n h(a)}.
\label{13}
\ee
\par
Next, we define the Hubble parameter,
$H\equiv {\dot a}/a$,\footnote{Dots will always denote
differentiation with respect to $\tau$.} in the usual way.
With this definition,  Eq.(\ref{11}) 
can  be re-expressed in the following form: 
\be
H^2={k\over a^2}-{1\over L^2} +{\omega_{n+1}M\over a^{n+1}}
+{\sigma^2\over n^2},
\label{14}
\ee
where we have also applied Eqs.(\ref{2},\ref{13}).
\par
In this model, the brane tension ($\sigma$) is
a free parameter that can be conveniently fine tuned.
Here, we choose
 $\sigma^2=n^2/L^2$ and thus cancel off 
the $a$-independent terms in Eq.(\ref{14}). This choice yields a 
(first) Friedmann-like equation:
\be
H^2={k\over a^2}+ {\omega_{n+1}M\over a^{n+1}}.
\label{15}
\ee
\par
Furthermore, we can take the $\tau$ derivative of the above equation,
which leads to
the associated  second Friedmann  equation:
\be
{\dot H}= -{k\over a^2}- 
{(n+1)\omega_{n+1}M\over 2  a^{n+1}}.
\label{16}
\ee
Note that the  TdS bulk  effectively induces  radiative matter
($\sim M/a^{n+1}$) 
in the brane universe.

\section{Euclidean CFT on the Brane}

Before proceeding, let us clarify  the underlying  premise of the 
dS/CFT correspondence.  It has been conjectured that 
the  thermodynamics
 of a dS cosmological horizon can  be directly associated
with the thermodynamics of a dual CFT.  Significantly, this CFT
 should be  a Euclidean one and living on an asymptotic boundary
 (in this case, the brane).
On the basis of such considerations, we will identify    
the brane (CFT) thermodynamics  by suitably adapting the AdS
analysis of Ref.\cite{sav}.
\par
We begin here by noting the following observation:  the metric
for a boundary CFT can only be determined up to
a conformal factor \cite{gub,wit}. Keeping this in mind,  let us 
  consider the asymptotic form of the  TdS metric:
\be
\lim_{a\rightarrow\infty}\left[{L^2\over a^2}ds^2_{n+2}\right]=
dt^2 + L^2d\Omega^2_n,
\label{17}
\ee
which can also  be identified with the Euclidean  metric for
the relevant  CFT.
  Evidently,
if the radius of the spatial sphere is to
 be set  equal to $a$,
 the Euclidean CFT time 
 must be rescaled by a factor of $a/L$.
 It follows 
 that the same factor ($a/L$) will
turn up when  the
thermodynamic properties of the dual spacetimes are related. (With one notable
exception:  the relation between the entropies \cite{wit}.)
\par
In view of the above discussion,  
the thermodynamic properties of the CFT can be expressed as follows \cite{sav}:
\be
E\equiv E_{CFT}={LM\over a},
\label{18}
\ee
\bea
T\equiv T_{CFT}&=&{L\over a}T_{dS}
\nonumber \\
&=& {1\over 4\pi a}\left[ {(n+1)a_{H}\over L}-{(n-1)Lk\over  a_H} \right],
\label{19}
\eea
\bea
S\equiv S_{CFT}&=&S_{dS} 
\nonumber \\
&=& {a_{H}^n V_{n}\over 4 G_{n+2}}.
\label{20}
\eea
\par
In relevance to the above, let us note the following.  The gravitational energy
 associated with this type of asymptotically dS geometry is always greater 
than that of the  pure (i.e., $M=0$) dS spacetime \cite{last4,new3}.
This is a reversal from the case of  Schwarzschild (and
Reissner-Nordstrom) dS   geometries, where a positive-mass
 black hole leads to an excitation of negative gravitational energy
\cite{baln,myun}. For this reason, 
 the CFT energy (\ref{18}) has  been
defined here as a positive quantity; in direct contrast
to a previous study \cite{med}.\footnote{It should be further noted
that, in Eq.(\ref{18}), we have omitted the energy associated
with the pure ($M=0$) dS background. This is consistent
with the convention initiated in Refs.\cite{ver,sav}.} 
\par
At this point, it is helpful to define  an energy density 
($\rho\equiv E/V$) and  pressure
($p\equiv \rho/n$);\footnote{Note that $p=\rho/n$ is the standard
equation of state for radiative matter.}
where
$V=a^nV_n$ is the volume of the brane universe.
\par
With the above definitions, the first and second 
 Friedmann-like 
equations (\ref{15},\ref{16})
can be re-expressed in the following form:
\be
H^2={16\pi G\over n(n-1)}\rho
+{k\over a^2},
\label{22}
\ee
\be
{\dot H}=-{8\pi G\over (n-1)}\left[\rho+p\right]-{k\over a^2}.
\label{23}
\ee
Here, we have used:
\be 
G={(n-1)\over L}G_{n+2},
\label{666}
\ee
where $G$ 
is the effective Newton constant on the
brane.\footnote{This  relation between bulk and brane gravitational constants
is the usual one for a Randall-Sundrum brane world (generalized
to arbitrary dimensionality) \cite{ran}.}
Significantly, the cosmological evolution
can now be directly attributed to 
 the energy density and pressure of radiative matter.
\par
For later convenience, we point out that the Friedmann
equations (\ref{22},\ref{23}) can alternatively be expressed  as follows:
\be
S_H={2\pi a\over n}\sqrt{E_{BH}\left[2E+kE_{BH}\right]},
\label{24}
\ee
\be
-kE_{BH}=n\left[E+pV -T_HS_H\right],
\label{25}
\ee
where we have defined:
\be
S_{H}\equiv (n-1) {HV\over 4G},
\label{26}
\ee
\be
E_{BH}\equiv n(n-1){V\over 8\pi G a^2},
\label{27}
\ee
\be
T_{H}\equiv -{{\dot H}\over 2\pi H}.
\label{28}
\ee
\par
The first Friedmann equation (Eq.(\ref{22}) or (\ref{24})) can 
also  be expressed in the following suggestive manner:
\be
S_{H}^2=2 S_B S_{BH}+k S_{BH}^2,
\label{29}
\ee
where:
\be
S_{B}\equiv {2\pi a\over n} E,
\label{30}
\ee
\be
S_{BH}\equiv {(n-1)\over 4 G a} V.
\label{31}
\ee
The parameters of
Eqs.(\ref{26}-\ref{28},\ref{30}-\ref{31})
are identical to  those  
defined in  Ref.\cite{ver}.
 For an AdS bulk, each of 
these parameters
 plays a significant role with regard to holographic bounds. 
(See Refs.\cite{ver,
sav} for a complete discussion.)  At this point,  we have introduced
the parameters  for illustrative   convenience and remind the reader that
 their respective roles  do  not necessarily
translate over to  a dS holographic theory.  
We consider this issue further in Section 5.
\par

\section{Thermodynamics at the Horizon and the Cardy-Verlinde Entropy}

One of the remarkable outcomes of  Ref.\cite{sav}
was the coincidence of two distinct theories at a special
moment in   the  evolution of the brane (in an AdS bulk).
In particular, it was demonstrated that
 the CFT thermodynamic relations  coincide 
 with the cosmological (i.e., Friedmann) equations 
when the brane crosses the black hole horizon.
Our current interest is  to ascertain if the same behavior occurs at the
cosmological horizon of a TdS bulk spacetime.
\par
We begin here by comparing  Eq.(\ref{15}) for $H^2$ with  the equation
for  the cosmological horizon (\ref{5}).  One can easily observe
 that  the Hubble constant 
must obey:
\be
H=\pm {1\over L}\quad\quad\quad at\quad a=a_{H}.
\label{33}
\ee
The $+$ sign indicates an expanding brane universe, while
the $-$ sign describes a brane universe that is contracting.
For illustrative purposes, we will subsequently
 focus on the expanding 
case.
\par
Next, let us reconsider  Eq.(\ref{20}) for the CFT entropy. 
As one might anticipate
(given  the second law of thermodynamics),
this total entropy remains constant as the system temporally evolves.
However, this is not true of  the entropy density:
\be
s\equiv {S\over V}= {(n-1)a_H^n\over 4G L a^n},
\label{34}
\ee 
which certainly  evolves  along with the radial size of the brane.
\par
When the brane crosses the horizon, this entropy density is given by:
\be
s={(n-1)H \over 4G}\quad\quad\quad at \quad a=a_H.
\label{35}
\ee
It directly follows  that (cf. Eq.(\ref{26})):
\be
S=S_H \quad\quad\quad at\quad a=a_H.
\label{36}
\ee
\par
It is of similar interest to consider the CFT temperature (\ref{19})
when the brane and horizon meet up.  By applying  Eq.(\ref{23}) for ${\dot H}$,
along with Eqs.(\ref{5},\ref{28},\ref{33}), we find that:
\be
T=-{{\dot H}\over 2\pi H}=T_H \quad\quad\quad at\quad a=a_H.
\label{37}
\ee
Hence, when the brane crosses the horizon, the CFT entropy and 
temperature can be
 simply expressed in terms of the Hubble parameter
and its derivative.  These expressions are universal
insofar as they do {\bf not} depend explicitly
on $M$ or $k$ (i.e., the parameters  describing  the TdS geometry).
\par
Let us now introduce a quantity that can be readily identified
with the Casimir energy of the brane universe \cite{ver,sav}:
\be
E_C\equiv n\left[E+pV-TS\right].
\label{38}
\ee
Given that $T=T_H$ and  $S=S_H$  at $a=a_H$,
we can further  deduce that (cf. Eq.(\ref{25})):
\be
E_C=n\left[E+pV-T_H S_H\right]=-kE_{BH} \quad\quad\quad at \quad a=a_H.
\label{39}
\ee
We will elaborate on the significance of  
the Casimir energy below.
\par
Let us now reconsider   the  scenario  of a generically
positioned  brane radius. As one might expect,
 the CFT thermodynamic properties can be shown to
satisfy the  first law of thermodynamics. That is:
\be
TdS=dE+PdV.
\label{40}
\ee
It is indeed more revealing
 if  the first law is reformulated  in terms of densities.
This expression takes on the following form:
\be
Tds=d\rho +n\left[\rho+p -Ts\right]{da\over a},
\label{41}
\ee
where we have applied the equation of state ($p=\rho/n$)
and $dV=nVda/a$ (since $V\sim a^n$). 
\par
As expressed above,
the  square-bracket combination  represents 
 the sub-extensive contribution to the thermodynamic
system.  Such a contribution should effectively describe the Casimir energy,
which notably agrees with our prior definition 
(\ref{38}).
Next, we will obtain  a more explicit form of  this Casimir
contribution.
\par
As an initial  step in this process, it is helpful if
 the CFT energy density is re-expressed (by way of Eqs.(\ref{5},\ref{18})) 
as follows:
\bea 
\rho &=& {ML\over a^{n+1}V_n}
\nonumber
\\
&=& {na_H^n\over 16\pi G_{n+2}a^{n+1}}\left[{a_H\over L}-{k L\over a_H}
\right].
\label{42}
\eea
\par
Next, we incorporate $p=\rho/n$, Eq.(\ref{34}) for $s$ and  Eq.(\ref{19})
for $T$  into the above expression. This procedure
ultimately yields:
\be
n\left[\rho+p-Ts\right]=-{2 k\gamma\over a^2},
\label{43}
\ee
where we have defined:
\be
\gamma\equiv {n(n-1)a_H^{n-1}\over 16\pi G a^{n-1}}.
\label{44}
\ee
\par
Comparing with Eq.(\ref{38}), which defines the Casimir energy, we have:
\be
E_C=-{2 k V \gamma \over a^2}=- k{ n(n-1)V_n a_H^{n-1}\over 8 \pi G a}.
\label{45}
\ee
Notably, this expression does not  depend explicitly on  the mass
parameter, $M$; although it does depend on the TdS geometrical
parameter, $k$.
\par
The above  formalism can be used to relate the entropy density (\ref{34}) 
and  the ``Casimir quantity'' (i.e., $\gamma$).
After some straightforward manipulations, we obtain:
\be
s^2=\left({4\pi\over n}\right)^2
\gamma \left[\rho+{k \gamma\over
a^2}\right].
\label{46}
\ee
Significantly, this entropy formula has a Cardy-like form \cite{car}.
Moreover, the Casimir-related quantity
 ($\gamma$) assumes  the role of the Cardy ``central 
charge''.\footnote{In Cardy's
formalism \cite{car}, the central charge describes the multiplicity
of massless particle species. It is clear that such a quantity should be
directly related to the Casimir energy density, as we have found.}
\par
Let us now  reconsider the special cosmological  moment; that is, 
when the brane passes through the
cosmological
horizon.
 At this coincidence point, Eq.(\ref{46}) leads directly to
 the first Friedmann-like  equation (\ref{22}).
Similarly, the second Friedmann-like equation (\ref{23}) follows
 when $a=a_H$ is  imposed on Eq.(\ref{43}).
Hence, we have extended the key results of Ref.\cite{sav} for the
case of a TdS bulk.

\section{Cosmological Considerations}
\par
In this section, we   examine some of the cosmological
implications of the prior results.
First, it is useful to re-express  the generalized Cardy-Verlinde
formula (\ref{46})  in the following equivalent form:
\be
S=\sqrt{{2\pi a\over n}S_C\left[2E
-E_C\right]},
\label{47}
\ee
where we have suitably defined the following Casimir entropy
(in analogy with Refs.\cite{ver,yumy}):
\bea
S_C&\equiv& \left.{2\pi a\over n} E_{BH} \right|_{a=a_H}
\nonumber \\
&=& {(n-1)V_n a_H^{n-1}\over 4 G}.
\label{48}
\eea
\par
Note that $S_C$ is strictly non-negative and independent of
$k$. This is in stark contrast to the Casimir energy. In fact,
the two quantities are related as follows (cf. Eq.(\ref{45})):
\be
E_C = -k {n \over  2\pi a } S_C.
\label{777}
\ee
Given that $S_C$ has no explicit  dependence on $k$ 
(which  can be $+1$, $0$ or $-1$),
we prefer to think of the Casimir entropy as the ``fundamental'' quantity,
from which $E_C$ can be obtained via the above ``definition'' (\ref{777}).
It just so happens that this definition  for $E_C$ coincides
precisely with the prior one (\ref{38}).
\par
With regard to the Casimir entropy, it is  particularly significant that:
\be
S_C=S_{BH} \quad\quad\quad at \quad a=a_H,
\label{xxx}
\ee
where $S_{BH}$ is 
 the  Bekenstein-Hawking entropy  of Eq.(\ref{31}).
Recall that a similar (but not exact)  coincidence  was found  between
the Casimir energy and  
$E_{BH}$;
cf. Eq.(\ref{39}).
\par
It can be readily  shown that, when $a=a_H$,
 the total entropy ($S$) actually
coincides with  the ``Hubble entropy'' ($S_H$) of Eq.(\ref{26}).
To   illustrate this occurrence, 
let us first consider the following equivalent
form of Eq.(\ref{47}):
\be
S^2=2 S_B S_C + k S_C^2,
\label{49}
\ee
where  $S_B$ is the ``Bekenstein entropy'' of Eq.(\ref{30}). 
Comparing Eq.(\ref{49}) for $S$ with Eq.(\ref{29}) for $S_H$,
we clearly observe the  equivalence of these two entropies
when the brane crosses the horizon.
\par
Given the outcomes of the seminal studies \cite{ver,sav},
one might wonder if  such  entropic coincidences
(at $a=a_H$) actually represent the saturation points of
holographic bounds.  It turns out that this is indeed
the case, provided that  Verlinde's conjectured
upper bound on the Casimir entropy \cite{ver}:
\be
S_C \leq S_{BH}
\label{52}
\ee
continues to hold. Verlinde originally proposed this
 universal bound
on the premise of  a holographic
upper limit on  the degrees
of freedom of the CFT as measured by the Casimir entropy.
  There seems no reason that such a bound
would fail to persist in our model given the following points. 
 (i) $S_C$ and $S_{BH}$ are
equivalent  
 when the brane crosses the horizon.  (ii) The Casimir
entropy has  no explicit dependence on  
 $k$ or  $M$ and, hence,  is  not sensitive to the
details of the TdS geometry.  It is interesting to note
that this bound implies (cf. Eqs.(\ref{31},\ref{48}))
that $a\geq a_{H}$; that is, the brane must remain outside of
(or at) the horizon.
\par
Again taking our cue from Verlinde, 
let us now make the distinction between  a strongly and
weakly self-gravitating  brane universe.  In the prior work \cite{ver},
a strongly (weakly) self-gravitating regime  was defined
by the condition: $Ha\geq 1$  ($Ha\leq 1$).
With this definition, Verlinde was able to deduce the following  
 \cite{ver}:
\be
S_B\geq S_{BH}\quad\quad\quad and\quad\quad\quad E \geq E_{BH}
\quad\quad\quad for \quad Ha\geq 1,
\label{888}
\ee
\be
S_B\leq S_{BH}\quad\quad\quad and\quad\quad\quad E\leq E_{BH}
\quad\quad\quad for \quad Ha\leq 1.
\label{889}
\ee 
\par
As it so happens,  virtually the same set of criteria are obtainable
for the TdS-bulk model, with only a minor modification.
Incorporating Eq.(\ref{22}) for $H^2$ into the appropriate
defining relations (\ref{27},\ref{30},\ref{31}), we find: 
\be
S_B\geq S_{BH}\quad\quad\quad and\quad\quad\quad E \geq E_{BH}
\quad\quad\quad for \quad Ha\geq \sqrt{2+k},
\label{998}
\ee
\be
S_B\leq S_{BH}\quad\quad\quad and\quad\quad\quad E\leq E_{BH}
\quad\quad\quad for \quad Ha\leq \sqrt{2+k}.
\label{999}
\ee 
That is, the definition of a  strongly (weakly) self-gravitating universe
must now be revised to incorporate the value of $k$, but
the  general formalism otherwise persists. 
Note that it is the $k=-1$ (hyperbolic) case
that exactly  reproduces the original Schwarzschild-AdS criteria.
\par
Let us now investigate the possibility of holographic bounds on
the CFT total entropy, $S$.
First,  we consider a strongly self-gravitating regime, for which it follows
that    (cf. Eqs.(\ref{52},\ref{998})):
\be
 S_B\geq S_{BH}\geq S_C \quad\quad\quad for \quad   Ha\geq \sqrt{2+k}.
\label{333}
\ee 
It is clear from Eq.(\ref{49}), that $S$ is monotonically
increasing in $S_C$ (for any allowed $k$) as long as
$S_C\leq S_B$.  Also in evidence, $S$ will reach its
maximum value (for this range) when $S_C=S_B$. This means that,
for a strongly self-gravitating universe, $S$ will
reach its maximum value when $S_C=S_{BH}=S_{B}$.
Comparing Eq.(\ref{49}) with Eq.(\ref{29}) for the Hubble entropy
($S_H$), we now see that this maximum value of $S$ coincides
precisely  with $S_H$. 
 So, for  {\bf any} of the 
prescribed values of $k$,\footnote{In fact, hypothetically speaking,
 this bound would
remain valid for any $k\geq -2$.  For $k< -2$,
not only is the bound no longer valid, but the  
``litmus test'' ($Ha\geq\sqrt{2+k}$ versus $Ha\leq\sqrt{2+k}$)
clearly breaks down.} we can establish the following bound:
\be
S\leq S_{H} \quad\quad\quad  for \quad Ha\geq \sqrt{2+k}.
\label{555}
\ee
Hence, the $a=a_H$ coincidence of $S$ and $S_{H}$ 
can also  be viewed as the saturation point of a holographic bound.
\par
It is interesting to note that, by virtue of Eqs.(\ref{49},\ref{333}),
the condition $S_C\leq S$ follows automatically for
a  strongly self-gravitating universe (for any allowed $k$).
This bound  is intuitively expected, given that
a massive TdS solution induces a positive energy excitation on the brane.
This is a pleasant reversal from an analogous  study with regard
to ``conventional'' black hole-dS  solutions \cite{med}. In this prior
work, it was found that the total entropy is always bounded from
above by the Casimir contribution.
\par
Next, let us see what can be deduced for a weakly self-gravitating
universe.  It is instructive to begin with the $k=-1$ case,
for which Eq.(\ref{49}) takes on the form:
\be
S^2 + \left( S_B - S_C \right)^2 =S_B^2.
\label{weee}
\ee
If we accept  the intuitive bound of $S_C\leq S$ to be universally valid
  (see above),
 then the above relation further implies
that $S_B\geq S_C$.
It seems reasonable to assume that this  bound continues to  hold
 for any allowed value of $k$, and   we will proceed on this basis.
Using  this assumption, we know from above that   (for  any allowed $k$)
$S$ is  monotonically increasing in $S_C$ and reaches 
its maximum value when $S_C=S_B$.  Hence, Eq.(\ref{49})
implies the following bound for a weakly self-gravitating
universe:
\be
S\leq \sqrt{2+k}S_{B}\quad\quad\quad for \quad  Ha\leq \sqrt{2+k}.
\ee 
We again point out the necessity for an assumption
in establishing  this bound.  Hence, it  is on a somewhat weaker footing
than the rigorously confirmed  bound of  Eq.(\ref{555}).

\section{Conclusion}
\par

In the preceding paper, we have  considered a brane universe in a 
topological de Sitter background  spacetime. 
To begin the analysis, we  identified the thermodynamic properties
of the TdS cosmological horizon.  Brane
dynamics were subsequently  examined, and  it was 
demonstrated that (with a suitable choice of brane tension)
the  evolution equations  take on a Friedmann-like form.
\par
After these initial considerations, we applied the dS/CFT correspondence 
and  deduced the
thermodynamics  of  a Euclidean CFT that lives
on the brane.  We were then able to demonstrate that
the CFT thermodynamic properties coincide with the  
Friedmann-like equations  when the brane crosses
the cosmological horizon. Moreover,  it was shown that
 the CFT  entropy  can be expressed
in terms of  a generalized Cardy-Verlinde
formula \cite{car,ver}.
In this context, the Casimir energy (i.e., the sub-extensive energy
 contribution) adopts the role of the Cardy central
charge.
\par
Finally, some of the cosmological implications  of our results
were considered.
For instance, we found that  the Casimir entropy coincides
with the Bekenstein-Hawking entropy when the brane crosses the horizon.
A  similar coincidence was found  between the total entropy
and the so-called Hubble entropy.
On the basis of these results (and other considerations), we have conjectured
that an  upper bound  on the Casimir entropy persists
even for  the exotic topology of our model.  (Such a bound was
originally proposed by Verlinde \cite{ver}, for an AdS spacetime, as
a universal consequence of the holographic principle \cite{tho,sus}.)
With this conjecture, it thus follows (either directly
or indirectly) that the observed entropic coincidences 
actually represent the  saturation points of their respective  holographic
bounds.
\par
It is interesting to compare this   TdS bulk scenario with
the case of a   Schwarzschild-dS 
background spacetime. In a recent study on the latter 
\cite{med},\footnote{Ref.\cite{med} formally considered a 
Reissner-Nordstrom-dS background. However, it may be
trivially extended to the Schwarzschild-dS case 
 (i.e., vanishing electrostatic charge).} 
we identified several troublesome  issues: a negative energy
density on the brane,  the total brane entropy being
bounded from above by the Casimir contribution, and 
the  brane universe being constrained to
a weakly self-gravitating regime. Furthermore,
there remains  the open question  of  how to incorporate
the thermodynamics of the  black hole horizon
into the  proposed dS/CFT  duality (which, so far, only
seems to  probe  the cosmological horizon). However, 
as we have now shown, all of these issues can
be circumvented   by reversing  the sign of the mass
term (while maintaing a non-negative mass).
\par
Given the apparent resolution of the noted  issues,
the results of the current paper  seem to strengthen
the status of the dS/CFT correspondence.  And yet, such TdS
solutions have the unfortunate side-effect of a naked cosmological
singularity.  It remains a possibility,  however, that
there exists a well-defined CFT which
contains some appropriate  description  of the  TdS singularity.
In this event, the singular behavior in the bulk would not be problematic
from the perspective of a brane observer. Clearly,
this  point will require further investigation.
Thus, for the time being, the outcomes of this paper  should
be regarded as
speculative.

\section{Acknowledgments}
\par
The author  would like to thank  V.P.  Frolov  for helpful
conversations.

\par\vspace*{20pt}


\end{document}